\newcommand{\pcmq}{\mbox{cm$^{-2}$}}
\newcommand{\psec}{\mbox{s$^{-1}$}}

\newcommand{\funit}{\mbox{ph~\pcmq~\psec}}

\def\deg{\ensuremath{^\circ}}

\newcommand{\ra}{\mbox{$\alpha_{\rm J2000}$}}
\newcommand{\dec}{\mbox{$\delta_{\rm J2000}$}}
\newcommand{\hi}{\mbox{H\,{\scriptsize I}}}

\newcommand{\liso}{\mbox{$L_{\gamma}$}}
\newcommand{\lratio}{\mbox{$r_{\gamma}$}}
\newcommand{\flux}{\mbox{$F_{\gamma}$}}
\newcommand{\nh}{\mbox{$N_{\rm H}$}}

\newcommand{\emm}{\mbox{$\bar{q}_{\gamma}$}}
\newcommand{\Msol}{\mbox{${\rm M}_{\odot}$}}

\documentclass[longauth]{aa} 

\usepackage{graphicx}
\usepackage{txfonts}
\usepackage{natbib}
\usepackage{lineno}

\bibpunct{(}{)}{;}{a}{}{,} 
\begin{document}

\title{{\em Fermi} Large Area Telescope observations of Local Group galaxies: Detection of M31 and search for M33}
\titlerunning{{\em Fermi}/LAT observations of Local Group galaxies: Detection of M31 and search for M33}

\author{
A.~A.~Abdo$^{(1,2)}$ \and 
M.~Ackermann$^{(3)}$ \and 
M.~Ajello$^{(3)}$ \and 
A.~Allafort$^{(3)}$ \and 
W.~B.~Atwood$^{(4)}$ \and 
L.~Baldini$^{(5)}$ \and 
J.~Ballet$^{(6)}$ \and 
G.~Barbiellini$^{(7,8)}$ \and 
D.~Bastieri$^{(9,10)}$ \and 
K.~Bechtol$^{(3)}$
% ...
}
\authorrunning{LAT collaboration}

\institute{
\inst{1}~Space Science Division, Naval Research Laboratory, Washington, DC 20375, USA\\ 
\inst{2}~National Research Council Research Associate, National Academy of Sciences, Washington, DC 20001, USA\\ 
\inst{3}~W. W. Hansen Experimental Physics Laboratory, Kavli Institute for Particle Astrophysics and Cosmology, Department of Physics and SLAC National Accelerator Laboratory, Stanford University, Stanford, CA 94305, USA\\ 
\email{bechtol@stanford.edu} \\
\inst{4}~Santa Cruz Institute for Particle Physics, Department of Physics and Department of Astronomy and Astrophysics, University of California at Santa Cruz, Santa Cruz, CA 95064, USA\\ 
\inst{5}~Istituto Nazionale di Fisica Nucleare, Sezione di Pisa, I-56127 Pisa, Italy\\ 
\inst{6}~Laboratoire AIM, CEA-IRFU/CNRS/Universit\'e Paris Diderot, Service d'Astrophysique, CEA Saclay, 91191 Gif sur Yvette, France\\ 
\inst{7}~Istituto Nazionale di Fisica Nucleare, Sezione di Trieste, I-34127 Trieste, Italy\\ 
\inst{8}~Dipartimento di Fisica, Universit\`a di Trieste, I-34127 Trieste, Italy\\ 
\inst{9}~Istituto Nazionale di Fisica Nucleare, Sezione di Padova, I-35131 Padova, Italy\\ 
\inst{10}~Dipartimento di Fisica ``G. Galilei", Universit\`a di Padova, I-35131 Padova, Italy\\ 
\inst{11}~Istituto Nazionale di Fisica Nucleare, Sezione di Perugia, I-06123 Perugia, Italy\\ 
\inst{12}~Dipartimento di Fisica, Universit\`a degli Studi di Perugia, I-06123 Perugia, Italy\\ 
\inst{13}~Centre d'\'Etude Spatiale des Rayonnements, CNRS/UPS, BP 44346, F-31028 Toulouse Cedex 4, France\\ 
\email{knodlseder@cesr.fr} \\
\email{jean@cesr.fr} \\
\inst{14}~Department of Physics, Center for Cosmology and Astro-Particle Physics, The Ohio State University, Columbus, OH 43210, USA\\ 
\inst{15}~Dipartimento di Fisica ``M. Merlin" dell'Universit\`a e del Politecnico di Bari, I-70126 Bari, Italy\\ 
\inst{16}~Istituto Nazionale di Fisica Nucleare, Sezione di Bari, 70126 Bari, Italy\\ 
\inst{17}~Laboratoire Leprince-Ringuet, \'Ecole polytechnique, CNRS/IN2P3, Palaiseau, France\\ 
\inst{18}~Department of Physics, University of Washington, Seattle, WA 98195-1560, USA\\ 
\inst{19}~Institut de Ciencies de l'Espai (IEEC-CSIC), Campus UAB, 08193 Barcelona, Spain\\ 
\inst{20}~NASA Goddard Space Flight Center, Greenbelt, MD 20771, USA\\ 
\inst{21}~University College Dublin, Belfield, Dublin 4, Ireland\\ 
\inst{22}~INAF-Istituto di Astrofisica Spaziale e Fisica Cosmica, I-20133 Milano, Italy\\ 
\inst{23}~Center for Research and Exploration in Space Science and Technology (CRESST) and NASA Goddard Space Flight Center, Greenbelt, MD 20771, USA\\ 
\inst{24}~Department of Physics and Center for Space Sciences and Technology, University of Maryland Baltimore County, Baltimore, MD 21250, USA\\ 
\inst{25}~George Mason University, Fairfax, VA 22030, USA\\ 
\inst{26}~Laboratoire de Physique Th\'eorique et Astroparticules, Universit\'e Montpellier 2, CNRS/IN2P3, Montpellier, France\\ 
\inst{27}~Department of Physics, Stockholm University, AlbaNova, SE-106 91 Stockholm, Sweden\\ 
\inst{28}~The Oskar Klein Centre for Cosmoparticle Physics, AlbaNova, SE-106 91 Stockholm, Sweden\\ 
\inst{29}~Royal Swedish Academy of Sciences Research Fellow, funded by a grant from the K. A. Wallenberg Foundation\\ 
\inst{30}~Dipartimento di Fisica, Universit\`a di Udine and Istituto Nazionale di Fisica Nucleare, Sezione di Trieste, Gruppo Collegato di Udine, I-33100 Udine, Italy\\ 
\inst{31}~Osservatorio Astronomico di Trieste, Istituto Nazionale di Astrofisica, I-34143 Trieste, Italy\\ 
\inst{32}~Department of Physical Sciences, Hiroshima University, Higashi-Hiroshima, Hiroshima 739-8526, Japan\\ 
\inst{33}~INAF Istituto di Radioastronomia, 40129 Bologna, Italy\\ 
\inst{34}~Universit\'e Bordeaux 1, CNRS/IN2p3, Centre d'\'Etudes Nucl\'eaires de Bordeaux Gradignan, 33175 Gradignan, France\\ 
\inst{35}~Center for Space Plasma and Aeronomic Research (CSPAR), University of Alabama in Huntsville, Huntsville, AL 35899, USA\\ 
\inst{36}~Science Institute, University of Iceland, IS-107 Reykjavik, Iceland\\ 
\inst{37}~Department of Physics and Department of Astronomy, University of Maryland, College Park, MD 20742, USA\\ 
\inst{38}~Research Institute for Science and Engineering, Waseda University, 3-4-1, Okubo, Shinjuku, Tokyo, 169-8555 Japan\\ 
\inst{39}~Max-Planck Institut f\"ur extraterrestrische Physik, 85748 Garching, Germany\\ 
\email{martinp@mpe.mpg.de} \\
\inst{40}~Istituto Nazionale di Fisica Nucleare, Sezione di Roma ``Tor Vergata", I-00133 Roma, Italy\\ 
\inst{41}~Department of Physics and Astronomy, University of Denver, Denver, CO 80208, USA\\ 
\inst{42}~Hiroshima Astrophysical Science Center, Hiroshima University, Higashi-Hiroshima, Hiroshima 739-8526, Japan\\ 
\inst{43}~Institute of Space and Astronautical Science, JAXA, 3-1-1 Yoshinodai, Sagamihara, Kanagawa 229-8510, Japan\\ 
\inst{44}~Institut f\"ur Astro- und Teilchenphysik and Institut f\"ur Theoretische Physik, Leopold-Franzens-Universit\"at Innsbruck, A-6020 Innsbruck, Austria\\ 
\inst{45}~NYCB Real-Time Computing Inc., Lattingtown, NY 11560-1025, USA\\ 
\inst{46}~Department of Chemistry and Physics, Purdue University Calumet, Hammond, IN 46323-2094, USA\\ 
\inst{47}~Partially supported by the International Doctorate on Astroparticle Physics (IDAPP) program\\ 
\inst{48}~Instituci\'o Catalana de Recerca i Estudis Avan\c{c}ats (ICREA), Barcelona, Spain\\ 
\inst{49}~Consorzio Interuniversitario per la Fisica Spaziale (CIFS), I-10133 Torino, Italy\\ 
\inst{50}~INTEGRAL Science Data Centre, CH-1290 Versoix, Switzerland\\ 
\inst{51}~Dipartimento di Fisica, Universit\`a di Roma ``Tor Vergata", I-00133 Roma, Italy
}

\date{Received 15 September 2010 / Accepted 19 October 2010}

\abstract
%
{Cosmic rays (CRs) can be studied through the galaxy-wide gamma-ray emission that they 
generate when propagating in the interstellar medium. 
The comparison of the diffuse signals from different systems may inform us about the
key parameters in CR acceleration and transport.
}
%
{We aim to determine and compare the properties of the cosmic-ray-induced gamma-ray
emission of several Local Group galaxies.
}
%
{We use 2 years of nearly continuous sky-survey observations obtained with the Large Area 
Telescope aboard the \textit{Fermi} Gamma-ray Space Telescope to search for gamma-ray
emission from \object{M31} and \object{M33}.
We compare the results with those for the Large Magellanic Cloud, 
the Small Magellanic Cloud, the Milky Way, and the starburst galaxies \object{M82} and 
\object{NGC253}.
}
%
{We detect a gamma-ray signal at 5$\sigma$ significance in the energy
range 200 MeV -- 20 GeV that is consistent with originating from \object{M31}.
The integral photon flux above 100~MeV amounts to
$(9.1 \pm 1.9_{\rm stat} \pm 1.0_{\rm sys}) \times 10^{-9}$ \funit.
We find no evidence for emission from \object{M33} and derive an upper limit
on the photon flux $>100$~MeV of $5.1 \times 10^{-9}$ \funit\ ($2\sigma$).
Comparing these results to the properties of other Local Group galaxies, we find indications
of a correlation between star formation rate and gamma-ray luminosity that also holds for 
the starburst galaxies.}
%
{The gamma-ray luminosity of \object{M31} is about half that of the Milky Way, which implies
that the ratio between the average CR densities in  \object{M31} and the Milky Way amounts 
to $\xi = 0.35\pm0.25$.
The observed correlation between gamma-ray luminosity and star formation rate suggests 
that the flux of \object{M33} is not far below the current upper limit from the LAT observations.
}


\maketitle

\section{Introduction}
\label{sec:intro}

Cosmic rays (CRs) produce high-energy gamma rays through interactions with interstellar matter and radiation fields. 
The resulting diffuse emissions directly probe CR spectra and intensities in galactic environments \citep[e.g.][]{strong07}.
The detection of the Small Magellanic Cloud \citep[SMC;][]{abdo10a} and detailed studies of the Large Magellanic Cloud \citep[LMC;][]{abdo10b} and the Milky Way
\citep[MW; e.g.][]{abdo09a} with the data collected by the Large Area Telescope (LAT) onboard the {\em Fermi} Gamma-ray Space Telescope enable comparative
studies of cosmic rays in environments that differ in star formation rate (SFR), gas content, radiation fields, size, and metallicities.

Other galaxies in the Local Group that have been predicted to be detectable high-energy gamma-ray emitters are \object{M31} (Andromeda) and \object{M33} (Triangulum) due to their relatively high masses and proximity.
So far, neither of these galaxies has been convincingly detected in high-energy gamma rays.
\object{M31} was observed by SAS-2 \citep{fichtel75}, COS-B \citep{pollock81}, and EGRET \citep{sreekumar94}, with the most stringent upper limit provided by EGRET being $4.9 \times 10^{-8}$ \funit\ at a 95\% confidence level \citep[see Fig.~3 of][]{hartman99}. 
\object{M33} has also been observed by COS-B \citep{pollock81} and EGRET, providing an upper limit of $3.6 \times 10^{-8}$ \funit\ \citep[see Fig.~3 of][]{hartman99}.

By comparing M31 properties to those of the MW, \citet{ozel87} estimated that the ratio $\xi$ of the CR density in \object{M31} and in the MW is $\xi \simeq 1$ and 
computed an expected $>$100~MeV flux from \object{M31} of $2.4 \times 10^{-8} \xi$ \funit. 
\citet{pavlidou01} made a comparable prediction of $1 \times 10^{-8}$ \funit, based on the assumption that $\xi \approx 0.5$, which they derived by comparing the estimated supernova rate in \object{M31} and in the MW. 
Using the same approach, they also estimated the $>100$~MeV flux of \object{M33} to be $1.1 \times 10^{-9}$ \funit.

If these estimates are correct, \object{M31} should be detectable by the LAT after 2 years of sky survey observations, while \object{M33} still may fall below the current sensitivity limit.
In this letter we report our searches for gamma-ray emission from \object{M31} and \object{M33} with the LAT using almost 2 years of survey data.
While we detect for the first time \object{M31} just above the current sensitivity limit, we could only derive an upper limit for the flux from \object{M33}.

\section{Observations and analysis}
\label{sec:observation}

\subsection{Data selection and analysis methods}

The data used in this work have been acquired by the LAT  between 8 August 2008 and 30 July 2010, a period of 721 days during which the LAT scanned the sky nearly continuously.
Events satisfying the standard low-background event selection \citep[`Diffuse' events;][]{atwood09} and coming from zenith angles $<105\deg$ (to greatly reduce the contribution by Earth albedo gamma rays) were used.
Furthermore, we selected only events where the satellite rocking angle was less than $52\deg$.
We further restricted the analysis to photon energies above 200~MeV; below this energy, the effective area in the `Diffuse class' is relatively small and strongly dependent on energy.
All analysis was performed using the LAT Science Tools package, which is available from the Fermi Science Support Center.
Maximum likelihood analysis has been performed in binned mode using the tool {\tt gtlike}.
We used P6\_V3 post-launch instrument response functions that take into account pile-up and accidental coincidence effects in the detector subsystems.

\subsection{\object{M31}}
\label{sec:m31}

For the analysis of \object{M31} we selected all events within a rectangular region-of-interest (ROI) of size $10\deg \times 10\deg$ centred on 
$(\ra, \dec)=(00^{\rm h}42^{\rm m}44^{\rm s}, +41\deg16'09^{\prime\prime})$ and aligned in Galactic coordinates.
The gamma-ray background was modelled as a combination of diffuse model components
and 4 significant point sources\footnote{
  \object{1FGL J0102.2+4223}, 
  \object{1FGL J0105.7+3930},
  \object{1FGL J0023.0+4453} \citep{abdo10c},
  and a hard source (\mbox{$\Gamma \sim 1.7$}) located at 
  $(\ra, \dec)=(00^{\rm h}39^{\rm m}16^{\rm s}, +43\deg27'07^{\prime\prime})$.}
that we found within the ROI.
Galactic diffuse emission was modelled using an LAT collaboration internal update of the model {\tt gll\_iem\_v02} \citep[e.g.][]{abdo10c} refined by using an analysis of 21 months of LAT data and improved gas template maps with increased spatial resolution.
Particular care was taken to remove any contribution from \object{M31} and \object{M33} in the templates by excluding all gas with $V_{\rm LSR}<-50$~km s$^{-1}$ 
within $2\deg \times 3\deg$ wide boxes around $(l,b)=(121\deg,-21.5\deg)$ and $(l,b)=(133.5\deg,-31.5\deg)$ for \object{M31} and \object{M33}, respectively\footnote{
  For \object{M31}, the velocity cut left some residual in the \hi\ template owing to overlap in 
  velocity with the MW along one side of \object{M31}.}.
In contrast to {\tt gll\_iem\_v02}\footnote{
  See the Galactic diffuse model description at 
  http://fermi.gsfc.nasa.gov/ssc/data/access/lat/BackgroundModels.html.}, 
we did not include an E(B-V) template in the model because it includes some signal from these galaxies.
We verified that the omission of the E(B-V) template did not affect the global fit quality over the ROI.
The overall normalization of the Galactic diffuse emission has been left as a free parameter in the analysis.
The extragalactic and residual instrumental backgrounds were combined into a single component assumed to have an isotropic distribution and a power-law spectrum with free normalization and free spectral index.
The spectra of the 4 point sources were also modelled using power laws with free normalizations and free spectral indices.

\begin{figure*}
\centering
\includegraphics[width=9.1cm]{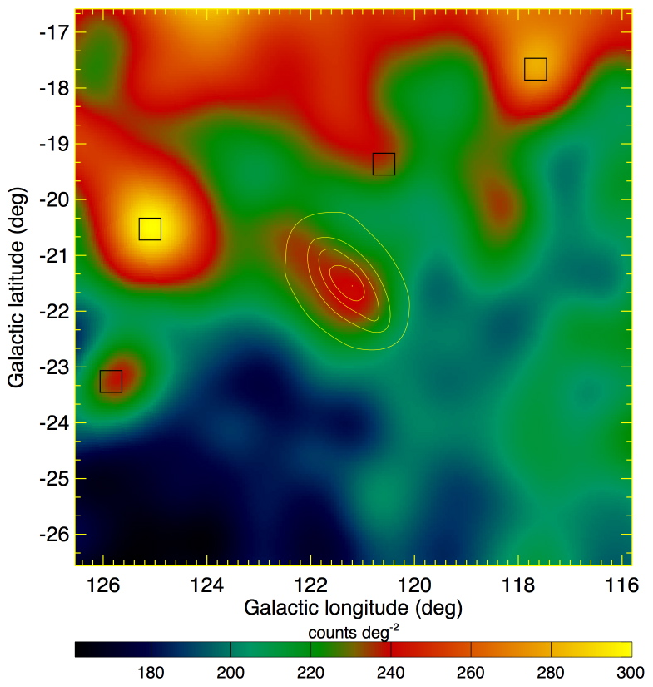}
\includegraphics[width=9.1cm]{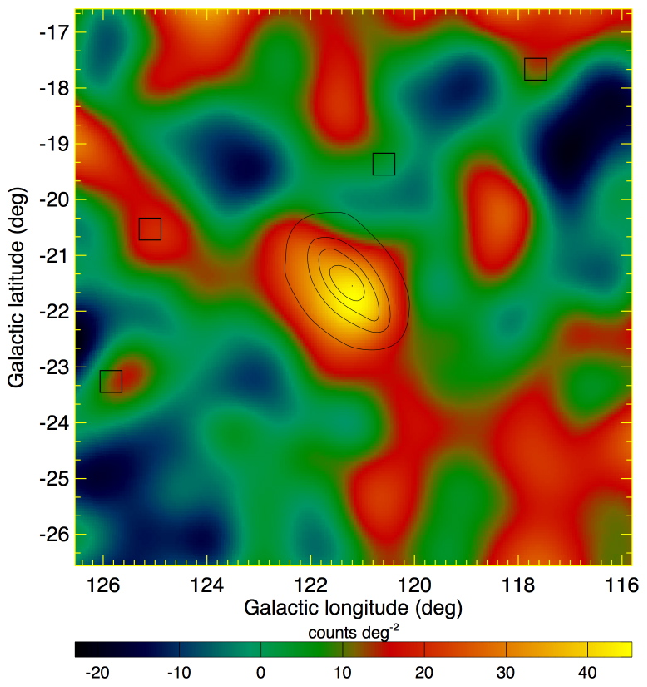}
\caption{
Gaussian kernel ($\sigma=0.5\deg$) smoothed counts maps of the region of interest (ROI) 
in a true local projection before (left) and after subtraction of the background model (right) 
for the energy range 200 MeV -- 20 GeV and for a pixel size of $0.05\deg \times 0.05\deg$.
Overlaid are IRIS 100 $\mu$m contours of \object{M31} convolved with the LAT point spread
function to indicate the extent and shape of the galaxy.
The boxes show the locations of the 4 point sources that have been included
in the background model.
\label{fig:image}
}
\end{figure*}

Figure~\ref{fig:image} shows LAT counts maps for the energy range 200 MeV - 20 GeV before (left panel) and after (right panel) subtracting the background model.
For the purpose of highlighting emission features on the angular scale of \object{M31}, the counts maps were smoothed using a 2D Gaussian kernel of $\sigma=0.5\deg$.
In this representation, an elongated feature that roughly follows the outline of \object{M31} (as indicated by black contours) is already visible in the counts map before background subtraction (left panel).
After this subtraction (right panel), this feature becomes the most prominent source of gamma-ray emission in the field.
The remainder of the structure in the `background subtracted' map is consistent with statistical fluctuations of the diffuse background emission, which illustrates that the signal from \object{M31} is close to the actual detection sensitivity of the LAT.

To test whether the emission feature is positionally consistent with \object{M31}, we performed maximum likelihood ratio tests for a grid of source positions centred
on the galaxy.
While the maximum likelihood ratio \citep[or the maximum {\em Test Statistic} value TS; cf.][]{mattox06} over the grid indicates the best-fitting source location, the decrease in TS from the maximum defines uncertainty contours that enclose the true source position at a given confidence level.
As usual, TS is defined as twice the difference between the log-likelihood of two alternative models $\mathcal{L}_1$ and $\mathcal{L}_0$, i.e. ${\rm TS} = 2(\mathcal{L}_1 - \mathcal{L}_0)$.
Using a point source with a power-law spectrum, we obtain a best-fitting location of $(\ra, \dec)=(00^{\rm h}42.4^{\rm m} \pm 1.4^{\rm m}, +41\deg10' \pm 11')$
for the gamma-ray source, which encloses the centre of \object{M31} within the $1\sigma$ confidence contour (quoted location uncertainties are at 95\% confidence).
Using instead of the point source an elliptically shaped uniform intensity region with a semi-major axis of $1.2\deg$, a semi-minor axis of $0.3\deg$ and a position
angle of $38\deg$ to approximate the extent and orientation of the galaxy on the sky\footnote{
  We estimated these parameters by adjusting an ellipse to the IRIS 100 $\mu$m map
  of \object{M31} \citep{miville05}.
},
we find a best-fitting location of $(\ra, \dec)=(00^{\rm h}43.9^{\rm m} \pm 1.8^{\rm m}, +41\deg23' \pm 22')$ that again encloses the centre of \object{M31} within the $1\sigma$ confidence contour.

We determined the statistical significance of the detection, as well as its spectral parameters, by fitting a spatial template for \object{M31} to the data on top of the 
gamma-ray background model that we introduced above. 
The \object{M31} template was derived from the Improved Reprocessing of the IRAS Survey (IRIS) 100 $\mu$m far infrared map \citep{miville05}.
Far infrared emission can be taken as a first-order approximation of the expected distribution of gamma-ray emission from a galaxy since it traces interstellar gas convolved with the recent massive star formation activity.
The spatial distributions of diffuse gamma-ray emission from our own Galaxy or the \object{LMC} are indeed traced by far-infrared emission to the first order.
From the IRIS 100 $\mu$m map, we removed any pedestal emission, which we estimated from an annulus around \object{M31}, and we clipped the image beyond a radius of $1.6\deg$.

Using this IRIS 100 $\mu$m spatial template for M31 and assuming a power-law spectral shape led to a detection above the background at TS~$=28.8$, which corresponds to a detection significance of $5.0\sigma$ for 2 free parameters.
We obtained a $>100$ MeV photon flux of $(11.0 \pm 4.7_{\rm stat} \pm 2.0_{\rm sys}) \times 10^{-9}$ \funit\
and a spectral index of $\Gamma=2.1 \pm 0.2_{\rm stat} \pm 0.1_{\rm sys}$
using this model.
Systematic errors include uncertainties in our knowledge of the effective area of the LAT and uncertainties in the modelling of diffuse Galactic gamma-ray emission.
As an alternative we fitted the data using the IRIS 60 $\mu$m, IRIS 25 $\mu$m, a template based on H$\alpha$ emission \citep{finkbeiner03} or the geometrical ellipse shape we used earlier for source localization.
All these templates provide results that are close to (and consistent with) those obtained using the IRIS 100 $\mu$m map.
Fitting the data using a point source at the centre of \object{M31} provided a slightly smaller TS ($25.5$) and a steeper spectral index ($\Gamma=2.5 \pm 0.2_{\rm stat} \pm 0.1_{\rm sys}$), which provides marginal evidence (at the $1.8\sigma$ confidence level) of a spatial extension of the source beyond the energy-dependent 
LAT point spread function.

Using the gamma-ray luminosity spectrum determined from a GALPROP model of the MW that was scaled to the assumed distance of $780$~kpc of \object{M31} \citep{strong10}\footnote{
  We use throughout this work a representative model of the MW from \citet{strong10}
  with a halo size of 4 kpc and that assumes diffusive reacceleration.
  The model is based on cosmic-ray, {\em Fermi}-LAT and other data, and includes 
  interstellar pion-decay, inverse Compton and Bremsstrahlung.
  Varying the halo size between 2 and 10 kpc affects the $>$100~MeV luminosity and
  photon flux by less than 10\% and 3\%, respectively.}
instead of a power law allows determination of the $>$100~MeV luminosity ratio \lratio\ between \object{M31} and the MW.
We obtain $\lratio = 0.55\pm0.11_{\rm stat}\pm0.10_{\rm sys}$ where we linearly added uncertainties in the assumed halo size of the model to the systematic 
errors in the measurement. The luminosity of \object{M31} is thus about half that of the MW. The model gives TS~$=28.9$, which is comparable to the value obtained using a power law, yet now with only one free parameter, the detection significance rises to $5.3\sigma$.
According to this model, the $>$100~MeV photon flux of \object{M31} is $(9.1 \pm 1.9_{\rm stat} \pm 1.0_{\rm sys}) \times 10^{-9}$ \funit.\\

\begin{figure}
\centering
\includegraphics[width=9.1cm]{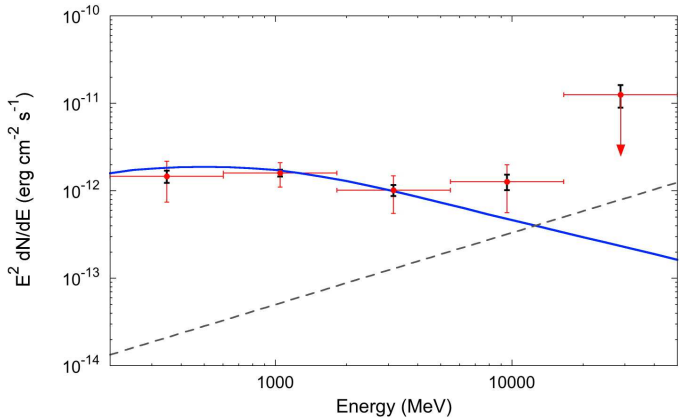}
\caption{
Spectrum of the \object{M31} emission obtained using the IRIS 100 $\mu$m spatial template.
Red error bars are statistical, black error bars are systematic uncertainties.
The solid line shows an MW gamma-ray luminosity model scaled to \object{M31} and the dashed 
one a possible contribution of \object{1ES 0037+405} (see text).
\label{fig:spectrum}
}
\end{figure}

\begin{table*}
\footnotesize
\caption{Properties and gamma-ray characteristics of Local Group and nearby starburst 
galaxies (see text).
\label{tab:properties}
}
\begin{center}
\begin{tabular}{lccccccc}
\hline
\hline
\noalign{\smallskip}
Galaxy & $d$ & $M_{\rm HI}$ & $M_{{\rm H}_2}$ & SFR & \flux\ & \liso\ & \emm\ \\
& kpc & 
$10^8$ \Msol & 
$10^8$ \Msol & 
\Msol\ yr$^{-1}$ & 
$10^{-8}$ \funit &
$10^{41}$ ph\,s$^{-1}$ &
$10^{-25}$ ph\,s$^{-1}$\,H-atom$^{-1}$ \\
\noalign{\smallskip}
\hline
\noalign{\smallskip}
MW & ... & 
   $35\pm4^{(7)}$ & $14\pm2^{(7)}$ & 
   $1-3^{(19)}$ & ... & $11.8 \pm 3.4^{(28)}$ &  $2.0\pm0.6$ \\
\object{M31} & $780\pm33^{(1)}$ & 
   $73\pm22^{(8)}$ & $3.6\pm1.8^{(14)}$ &
   $0.35-1^{(19)}$ & $0.9 \pm 0.2$ & $6.6\pm1.4$ & $0.7\pm0.3$ \\
\object{M33} & $847\pm60^{(2)}$ & 
   $19 \pm 8^{(9)}$ & $3.3 \pm 0.4^{(9)}$ &
   $0.26-0.7^{(20)}$ & $<0.5$ & $<5.0$ &  $<2.9$ \\
\object{LMC} & $50 \pm 2^{(3)}$ & 
  $4.8\pm0.2^{(10)}$ & $0.5\pm0.1^{(15)}$ &
  $0.20-0.25^{(21)}$ & $26.3\pm2.0^{(25)}$ & $0.78\pm0.08$ & $1.2\pm0.1$ \\
\object{SMC} & $61 \pm 3^{(4)}$ &
  $4.2\pm0.4^{(11)}$ & $0.25\pm0.15^{(16)}$ &
  $0.04-0.08^{(22)}$ & $3.7\pm0.7^{(26)}$ & $0.16\pm0.04$ & $0.31\pm0.07$ \\
\object{M82} & $3630 \pm 340^{(5)}$ & 
  $8.8\pm2.9^{(12)}$ & $5\pm4^{(17)}$ &
  $13-33^{(23)}$ & $1.6\pm0.5^{(27)}$ & $252\pm91$ & $158\pm75$ \\
\object{NGC253} & $3940 \pm 370^{(6)}$ & 
  $64\pm14^{(13)}$ & $40\pm8^{(18)}$ &
  $3.5-10.4^{(24)}$ & $0.6\pm0.4^{(27)}$ & $112\pm78$ & $9\pm6$ \\
\noalign{\smallskip}
\hline
\end{tabular}
\end{center}
{\bf References.}
(1) \citet{stanek98}; 
(2) \citet{galleti04}; 
(3) \citet{pietrzynski09}; 
(4) \citet{hilditch05}; 
(5) \citet{karachentsev02}; 
(6) \citet{karachentsev03}; 
(7) \citet{paladini07}; 
(8) \citet{braun09}; 
(9) \citet{gratier10}; 
(10) \citet{staveley03}; 
(11) \citet{stanimirovic99}; 
(12) \citet{chynoweth08}; 
(13) \citet{combes77}; 
(14) \citet{nieten06}; 
(15) \citet{fukui08}; 
(16) \citet{leroy07}; 
(17) \citet{mao00}; 
(18) \citet{houghton97}; 
(19) \citet{yin09}; 
(20) \citet{gardan07}; 
(21) \citet{hughes07}; 
(22) \citet{wilke04}; 
(23) \citet{forster03}; 
(24) \citet{lenc06}; 
(25) \citet{abdo10b}; 
(26) \citet{abdo10a}; 
(27) \citet{abdo10d}; 
(28) \citet{strong10}: range based on GALPROP models with various halo sizes. 
\end{table*}
\indent We determined the spectrum of the gamma-ray emission from \object{M31} independently of any assumption about the spectral shape by fitting the IRIS 100 $\mu$m template in five logarithmically spaced energy bins covering the energy range 200 MeV -- 50 GeV to the data.
Figure \ref{fig:spectrum} shows the resulting spectrum on which we superimposed the GALPROP model of the MW for $\lratio = 0.55$.
Overall, the agreement between the observed spectrum of \object{M31} and the model is very satisfactory.
The upturn in the spectrum at high energies, though not significant, could possibly be attributed to emission from the BL Lac object \object{1ES 0037+405}, the only known blazar in the line of sight towards \object{M31}.
In a dedicated analysis above 5 GeV, we found a cluster of 6--7 counts that are positionally consistent with coming from that blazar. Adding \object{1ES 0037+405} as a point source to our model and extending the energy range for the fit to 200 MeV -- 300 GeV results in a TS$=16-20$ for the source, where the range reflects uncertainties in modelling the spectrum of the isotropic background component at energies $>$100~GeV. The fit suggests a hard power-law spectral index ($\Gamma=1.2 \pm 0.4$), which explains why the source is only seen at high energies. Within 200 MeV -- 20 GeV, however, the source contributes only $\sim$8 counts, a number that is tiny compared to the $\sim$240 counts that are attributed to \object{M31}. The impact of \object{1ES 0037+405} on the flux and gamma-ray luminosity estimates for \object{M31} is thus negligible.

We also repeated our analysis for a larger ROI of size $20\deg \times 20\deg$ in which we found 14 point sources in our LAT internal source list.
Searching for the faint signal from \object{M31} in such a large ROI relies on the accurate modelling of the spatial distribution of the diffuse gamma-ray background over a large area, which is an important potential source of systematic uncertainties.
Nevertheless, results obtained for this large ROI were consistent with those obtained for the $10\deg \times 10\deg$ ROI.

\subsection{\object{M33}}
\label{sec:m33}

For the analysis of \object{M33} we selected all events within a rectangular ROI of size $10\deg \times 10\deg$ centred on $(\ra, \dec)=(01^{\rm h}33^{\rm m}51^{\rm s}, +30\deg39'37^{\prime\prime})$ and aligned in Galactic coordinates.
Within this field we detected 3 background point sources\footnote{
  \object{1FGL J0134.4+2632}, 
  \object{1FGL J0144.6+2703}, and
  \object{1FGL J0112.9+3207} \citep{abdo10c}.}
that we included in the background model.
The remainder of the analysis was similar to what was done for \object{M31}.

We did not detect any significant signal towards the direction of \object{M33}.
Using a spatial template based on the IRIS 100 $\mu$m map of \object{M33} and taking the GALPROP models of \citet{strong10} for the spectral shape, we derived an upper $>$100~MeV flux limit of $5.1 \times 10^{-9}$ \funit\ ($2\sigma$).

\section{Discussion}
\label{sec:discussion}

Based on the flux \flux\ measured for \object{M31} and the flux upper limit for \object{M33}, we computed the $>$100~MeV photon luminosities
$\liso = 4 \pi d^2 \flux$ and average emissivities $\emm = \liso / \nh$, which we compare to the values obtained for the MW, the \object{LMC}, and the \object{SMC} 
(see Table \ref{tab:properties}). 
Here, $d$ is the distance of the galaxy and $\nh = 1.19 \times 10^{57} (M_{\rm HI} + M_{{\rm H}_2})$ is the total number of hydrogen atoms in a galaxy, with $M_{\rm HI}$ and $M_{{\rm H}_2}$ in units of \Msol.
Quoted uncertainties in \liso\ and \emm\ include uncertainties in distance and hydrogen mass of the galaxies.
The variations in \liso\ and \emm\ from one galaxy to another may inform us about how the CR population is affected by global galactic properties. 
From the \emm\ values, we estimate the ratio $\xi$ of the average CR density in \object{M31} and in the MW to $\xi=0.35\pm0.25$, consistent with the estimate of \citet{pavlidou01}. 
On the other hand, the flux upper limit for \object{M33} allows for an average CR density in that galaxy that is above the MW value, hence up to a few times greater than the $\xi=0.2$ estimated by \citet{pavlidou01}.

\begin{figure}
\centering
\includegraphics[width=\columnwidth]{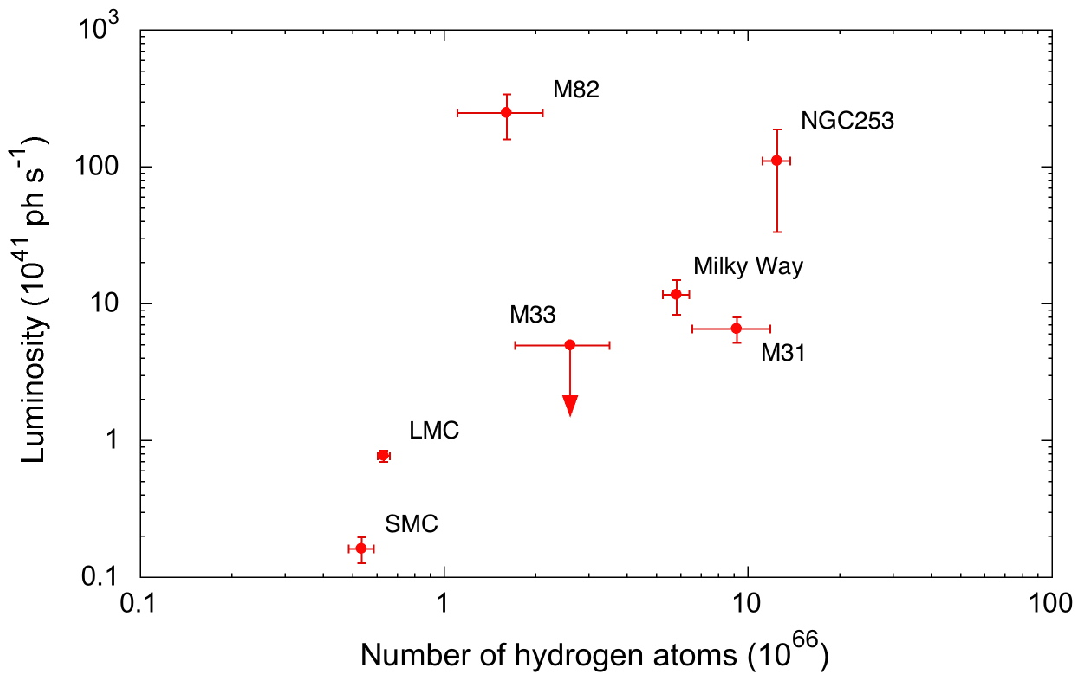}
\includegraphics[width=\columnwidth]{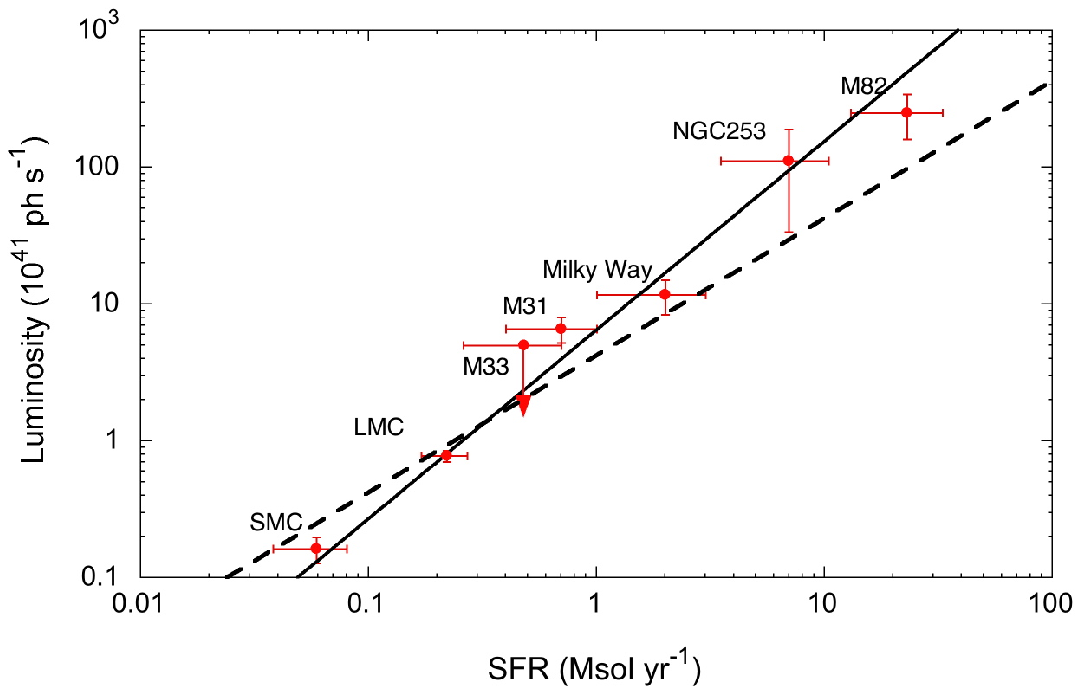}
\caption{
Gamma-ray $>100$~MeV luminosity versus total number of hydrogen atoms (top panel) and star formation rate (bottom panel) for Local Group galaxies and the starbursts \object{M82} and \object{NGC253}.
In the bottom panel, the lines are power-law fits to the data for the MW, \object{M31}, the 
\object{LMC}, and the \object{SMC}, for which the slope was free (solid) or fixed to 1 (dashed).
\label{fig:lsfr}
}
\end{figure}

By comparing the \liso\ of our sample of Local Group galaxies to their total hydrogen masses and SFRs, we find a close correlation between \liso\ and SFR and greater scatter between \liso\ and gas mass (see Fig.~\ref{fig:lsfr}). In the bottom panel of Fig.~\ref{fig:lsfr}, the ranges of SFR values, which have been rescaled to the distances $d$ adopted here, reflect uncertainties in the SFR estimates based on the various methods used to determine them (see Table \ref{tab:properties}).
There is a clear trend toward increasing \liso\ with increasing SFR, with $\liso = (7.4 \pm 1.6) \times {\rm SFR}^{1.4\pm0.3}$
when fitted by a power law, where \liso\ and SFR are in units of $10^{41}$ ph\,s$^{-1}$ and \Msol\,yr$^{-1}$, respectively.
We also added the luminosities derived by \citet{abdo10d} for \object{M82} and \object{NGC253} to this plot, illustrating that the relation obtained for Local Group galaxies also holds for nearby starburst galaxies. 
Assuming that it also holds for \object{M33} allows estimation of the luminosity of $\liso \sim (1-4) \times 10^{41}$ ph\,s$^{-1}$ for this galaxy, corresponding to a $>$100~MeV flux of $(1-4) \times 10^{-9}$ \funit. \object{M33} thus may be within reach of the LAT within the next few years.

The \liso-SFR plot does suggest a correlation in common for Local Group and starburst galaxies. Although it is premature to draw conclusions about any strong correlation over such a wide range of galaxy properties because of the small size of our sample, if such a correlation exists, it would be analogous to the well-known tight correlation between radio and far-infrared emission over a wide range of galaxy types \citep[e.g.][]{murphy06}. The latter is linked to the relation between CRs and SFR, and although not yet fully understood, it is thought to result to some extent from CR electron calorimetry. While proton calorimetry clearly can be excluded as an explanation of the \liso-SFR correlation because the intermediate-size galaxies of the Local Group are thought to be very inefficient at retaining CR protons, the dominant CR component \citep{strong10}, a correlation may relate to the contribution of CR leptons to the gamma-ray emission. Depending on the ISM and CR transport conditions, CR leptons may lose their energy predominantly through gamma-ray-emitting processes (like inverse-Compton or Bremsstrahlung, as opposed to ionization and synchrotron) and dominate the total gamma-ray luminosity\footnote{
  Some variants of the GALPROP MW model actually predict that 
  leptons can be responsible for up to $\sim$50\% of its $>$100~MeV gamma-ray 
  photon flux \citep{strong10}.
}. This could drive the correlation between \liso\ and SFR for galactic systems with high lepton calorimetric efficiency. Whatever the explanation for this global correlation, it is worthwhile noting that it holds despite the fact that conditions may vary considerably within a galaxy (e.g. the peculiar 30 Doradus region in the LMC, or the very active cores of starbursts).

The \liso\ vs SFR plane therefore seems to hold potential for defining constraints on CR production and transport processes. The inferred \liso\ values are, however, not uniquely due to CR-ISM interactions but include a contribution of individual galactic sources such as pulsars and their nebulae. The relative contributions of discrete sources and CR-ISM interactions to the total gamma-ray emission very likely vary with galaxy properties like SFR, which may complicate the interpretation of any \liso\ trend in terms of CR large-scale population and transport.

Also more exotic processes, such as annihilation or decay of WIMPs (weakly interacting massive particles), might contribute to the overall signal from \object{M31}. Several extensions of the Standard Model of particle physics naturally predict the existence of WIMPs (e.g. supersymmetry, universal extra dimensions). Rather than focusing on a specific scenario, we estimate a conservative upper bound on this contribution in the case of a generic 100~GeV WIMP annihilating exclusively into bottom quarks, which is one of the leading tree level annihilation channels of a WIMP predicted by supersymmetric theories.
The normalization of the predicted spectrum is initially set to zero and is increased until it just meets, but does not exceed, the 95\% confidence upper limit on the measured \object{M31} spectrum at any energy. We find that when assuming an Einasto dark matter halo profile \citep{navarro10} that matches the \object{M31} kinematic data \citep{klypin02}, this contribution corresponds to a 95\% confidence upper limit on the annihilation cross section of approximately 5 $\times$ 10$^{-25}$ cm$^{3}$\,s$^{-1}$.

\begin{acknowledgements}
The \textit{Fermi} LAT Collaboration acknowledges support from a number of agencies
and institutes for both development and the operation of the LAT, as well as for scientific 
data analysis. 
These include NASA and DOE in the United States, CEA/Irfu and IN2P3/CNRS in France, 
ASI and INFN in Italy, MEXT, KEK, and JAXA in Japan, and the K.~A.~Wallenberg 
Foundation, the Swedish Research Council, and the National Space Board in Sweden. 
Additional support from INAF in Italy and CNES in France for science analysis during the 
operations phase is also gratefully acknowledged.
\end{acknowledgements}


\Online
\begin{appendix}

\section{Gamma-ray spectrum of \object{M31}}
\label{sec:spectrum}

Table \ref{tab:spectrum} provides the intensity values of the \object{M31} gamma-ray
spectrum that is shown in Fig.~\ref{fig:spectrum}.
Statistical errors are at the $1\sigma$ confidence level, and the upper limit for the 
16.6 -- 50.0 GeV energy bin at the $2\sigma$ confidence level.
Systematic errors include uncertainties in our knowledge of the effective area of 
the LAT and uncertainties in the modelling of diffuse Galactic 
gamma-ray emission.
The former were determined using modifications of the instrument response
functions that bracket the uncertainties in our knowledge of the LAT effective area.
The latter were determined by deriving spectra for variations of the diffuse Galactic models
that make use of either an E(B-V) template or for which the gas templates have been replaced
by the IRIS 100 $\mu$m map, from which emission associated to \object{M31} has been
removed.
Both types of systematic uncertainties were added linearily.

The last column gives the number of counts attributed to \object{M31} in each of the energy bins from the fit of a spatial model to the
present data.

\begin{table}[!h]
\caption{Measured spectrum of \object{M31} (see text).}
\label{tab:spectrum}
\begin{center}
\begin{tabular}{lcccc}
\hline
\hline
\noalign{\smallskip}
Energy & 
Intensity & 
Stat. error & 
Sys. error &
Counts \\
MeV & \multicolumn{3}{c}{$10^{-12}$ erg\,cm$^{-2}$\,s$^{-1}$} & \\
\noalign{\smallskip}
\hline
\noalign{\smallskip}
200 -- 603 & 1.46 & 0.71 & 0.23 & 118.9 \\
603 -- 1821 & 1.60 & 0.49 & 0.14 & 69.5 \\
1821 -- 5493 & 1.02 & 0.47 & 0.15 & 15.9 \\
5493 --16572 & 1.27 & 0.71 & 0.26 & 7.0 \\
16572 -- 50000 & $<12.5$ & ... & 3.6 & $<24.4$ \\
\noalign{\smallskip}
\hline
\end{tabular}
\end{center}
\end{table}

\end{appendix}


\end{document}